\documentclass[12pt,a4paper,twoside]{article}
\usepackage{epsf}
\usepackage{dcolumn}
\usepackage{rotate}

\usepackage{physics_symbols}

\newcommand{\ecmA}{191.6}
\newcommand{\ecmB}{195.6}
\newcommand{\lA}{29.0}
\newcommand{\lB}{69.5}

\newcommand{\smlim}{98.8}
\newcommand{\expsmlim}{99.9}

\newcommand{\cutlim}{97.6}
\newcommand{\cutexplim}{98.7}

\newcommand{\hlim}{85.2}
\newcommand{\exphlim}{86.1}

\newcommand{\tanb}{\ensuremath{\tan \beta}}
\newcommand{\sba}{\ensuremath{\sin^2(\beta-\alpha)}}

\newcommand{\nsig}{\ensuremath{N_{\rm sig}}}
\newcommand{\nbkg}{\ensuremath{N_{\rm bkg}}}
\newcommand{\nobs}{\ensuremath{N_{\rm obs}}}

\newcommand{\SM}{Standard Model\@}

\newcommand{\ALEPH}{{\sc aleph}}

\newcommand{\mh}{\ensuremath{m_{\h}}}
\newcommand{\mA}{\ensuremath{m_{\rm A}}}

\newcommand{\h}{\ensuremath{{\mathrm{h}}}}

\newcommand{\nn}{\ensuremath{\nu\overline{\nu}}}
\newcommand{\qq}{\ensuremath{\mathrm{q}\overline{\mathrm{q}}}}

\newcommand{\bbbar}{\ensuremath{\mathrm{b}\overline{\mathrm{b}}}}

\newcommand{\ee}{\ensuremath{{\mathrm{e}^+\mathrm{e}^-}}}
\newcommand{\lplm}{\ensuremath{{\ell^+\ell^-}}}

\newcommand{\tautau}{\ensuremath{\tau\tau}}

\newcommand{\bbbb}{\bbbar\bbbar}

\newcommand{\ttqq}{\ensuremath{\tautau\qq}}

\newcolumntype{d}[1]{D{.}{.}{#1}}

\begin{document}
 
\begin{titlepage}
\begin{center}
{\Large P R E L I M I N A R Y }
\end{center}
\begin{flushright}
4 August 1999
\end{flushright}
\vfill
\begin{center}

{\boldmath {\huge\bf Search for neutral Higgs bosons}\\
\mbox{}\\
{\huge\bf in \ee\ collisions at \roots\hspace*{5pt} $\leq$ 196 GeV}\\ \unboldmath}

\vfill
\mbox{}\\
{\large The ALEPH Collaboration}\\
Contact person: Jason Nielsen (Jason.Nielsen@cern.ch)\\

\mbox{}\\
\vfill

{\bf Abstract}
\end{center}

\noindent

A preliminary search for neutral Higgs bosons is performed in the data
collected by ALEPH in 1999, corresponding to integrated luminosities
of \lA\ and \lB~\invpb\ at centre-of-mass energies of \ecmA\ and
\ecmB~\gev\ respectively. No evidence for a signal is found. Combined
with the lower energy ALEPH data, this observation leads to a 95\%
confidence level lower mass limit of \smlim~\gevcc\ for the \SM\ Higgs
boson.  In the MSSM, the lower limit for the mass of the lightest
CP-even higgs boson h, is found to be \hlim~\gevcc\ for all values of
\tanb$\ge 1$.


\vfill
\centerline{{\it ALEPH contribution to the XIX International Symposium}}
\centerline{{\it on Lepton and Photon Interactions at High Energies}} 
\vfill\vfill
\end{titlepage}

\setcounter{page}{1}
 
\section{Summary of preliminary results}

The analyses designed for the 189~\gev\ data, described in detail 
in~\cite{higgs}, have been applied without any modifications
to the data collected by ALEPH up to July 27, 1999.  The data sample 
comprises \lA~\invpb\ at \roots=\ecmA~\gev\ and \lB~\invpb\ at 
\roots=\ecmB~\gev.

The expected numbers of background events, the expected signal for a
Higgs boson with \mh=100~\gevcc\ and the numbers of selected
candidates are listed in Table~\ref{tab:sumhZ} for the hZ analyses and
in Table~\ref{tab:sumhA} for the hA analyses. The reconstructed mass
of the candidates selected by the NN- and cut-based hZ searches are
displayed in Fig.~1(a) and (b).  No departure from the \SM\ expectations is
observed.  This leads to improved 95\% confidence level limits when
the latest data are combined with those collected by ALEPH at lower
energies. The confidence levels for the data in the background only
hypothesis or the background and signal hypothesis are shown in
Fig.~2(a) and (b).
The limit obtained for the \SM\ Higgs boson is
\mh$\ge$\smlim~\gevcc, with an expected limit of \expsmlim~\gevcc.
The limit is calculated as described in~\cite{higgs}, with the proper
treatment of the overlap between the hZ and hA four jet and
$\tau^+\tau^-$ selections. The stand-alone hZ observed limit from the
cut-based set of analyses is \cutlim~\gevcc, with \cutexplim~\gevcc\ 
for the expected limit. For the CP-even boson h of the MSSM, the limit
is \hlim~\gevcc, for all values of \tanb\ above 1, with
\exphlim~\gevcc\ expected.  Fig.~3 shows the excluded domain in the
\mh--\tanb plane.


\begin{table}[hbt]
\begin{center}
\begin{tabular}{|l|c|c|c|}
\hline\hline 
Channel            & \nbkg  &  \nsig  & \nobs \\
\hline 
h\qq\  (NN/cut)    &   15.0/12.4  &   4.6/3.9   &   12/9   \\ 
h\nn\  (NN/cut)    &   4.4/4.9  &   1.8/1.4   &   6/8   \\
h\lplm             &   10.4      &   0.8       &   7     \\
h\tautau\ + \ttqq\ &   2.5      &   0.4       &   2     \\
\hline
Total (NN/cut)    &  32.3/30.3 &   7.6/6.6   &  27/26  \\
\hline \hline 

\end{tabular}
\caption{ Number of background events expected, expected signal for 
\mh=100~\gevcc\ and observed candidates in the 1999 ALEPH data for
searches by cut- and NN-based sets of selections.
\label{tab:sumhZ}}
\end{center}
\end{table}

\begin{table}[hbt]
\begin{center}
\begin{tabular}{|l|c|c|c|}
\hline\hline 
Channel            & \nbkg  &  \nsig  & \nobs \\
\hline 
\bbbb              &  2.9   &   2.3   &   1   \\
\bbbar\tautau      &  1.9   &   0.4   &   0   \\
\hline
Total              &  4.8   &   2.7   &   1   \\
\hline \hline 

\end{tabular}
\caption{ Number of background events expected, expected signal for 
\mh $\approx$ \mA = 85~\gevcc\ and observed candidates in the 1999 ALEPH 
data with the hA selections.
\label{tab:sumhA}}
\end{center}
\end{table}

\begin{figure}
\begin{center}
\begin{picture}(180,80)
\put(-15,-5){\epsfxsize100mm\epsfbox{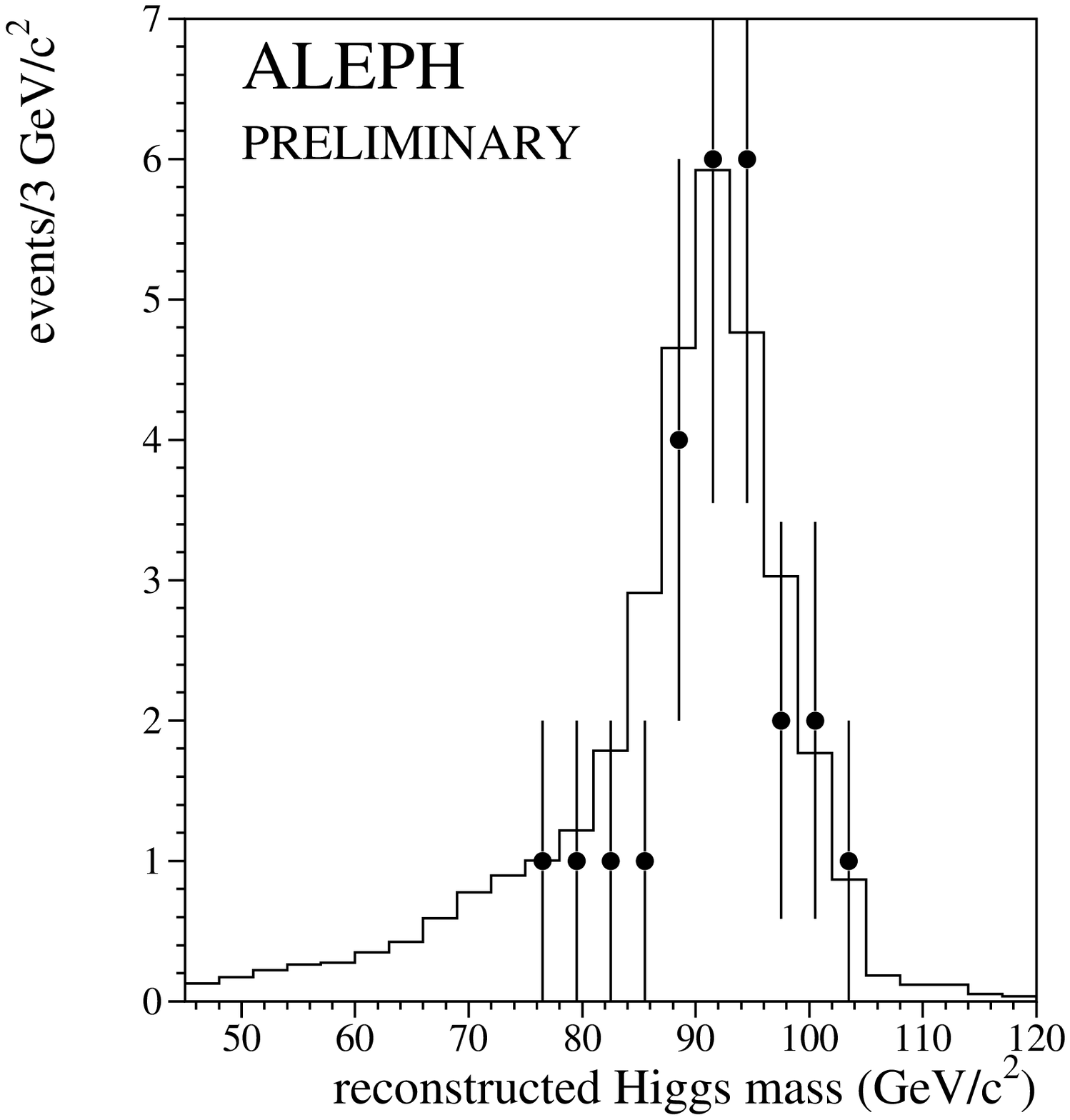}}
\put(70,-5){\epsfxsize100mm\epsfbox{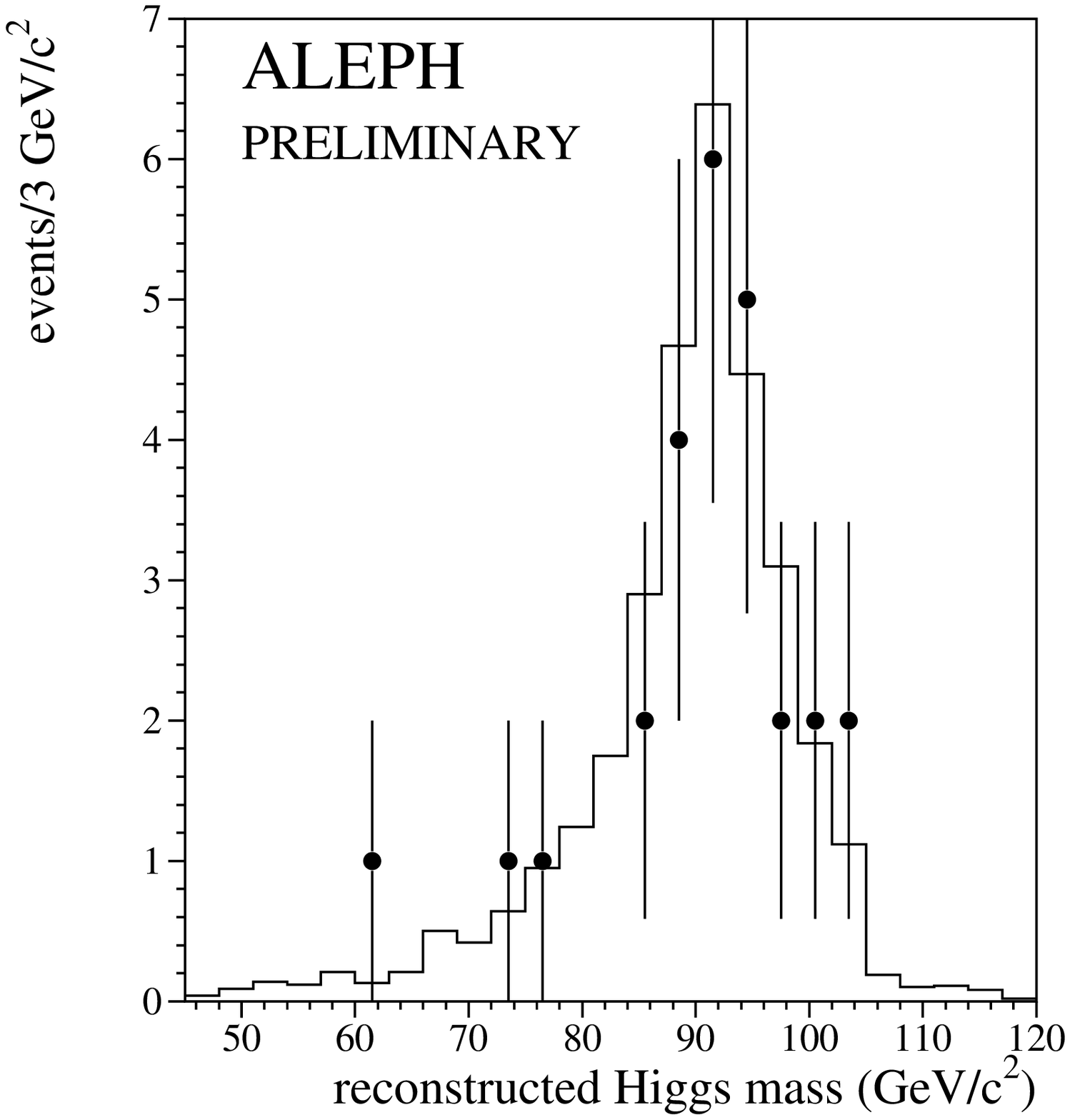}}
\put(10,68){\Large(a)}
\put(95,68){\Large(b)}
\end{picture}
\caption{ Distribution of the reconstructed Higgs boson mass of 
the selected events (dots) in the hZ searches by the NN-based (a) and
cut-based (b) sets of selections. The histograms show the Standard Model 
expectations. 
\label{fig:mass_19x}} 
\end{center}
\end{figure}

\begin{figure}
\begin{center}
\begin{picture}(180,80)
\put(-15,-5){\epsfxsize100mm\epsfbox{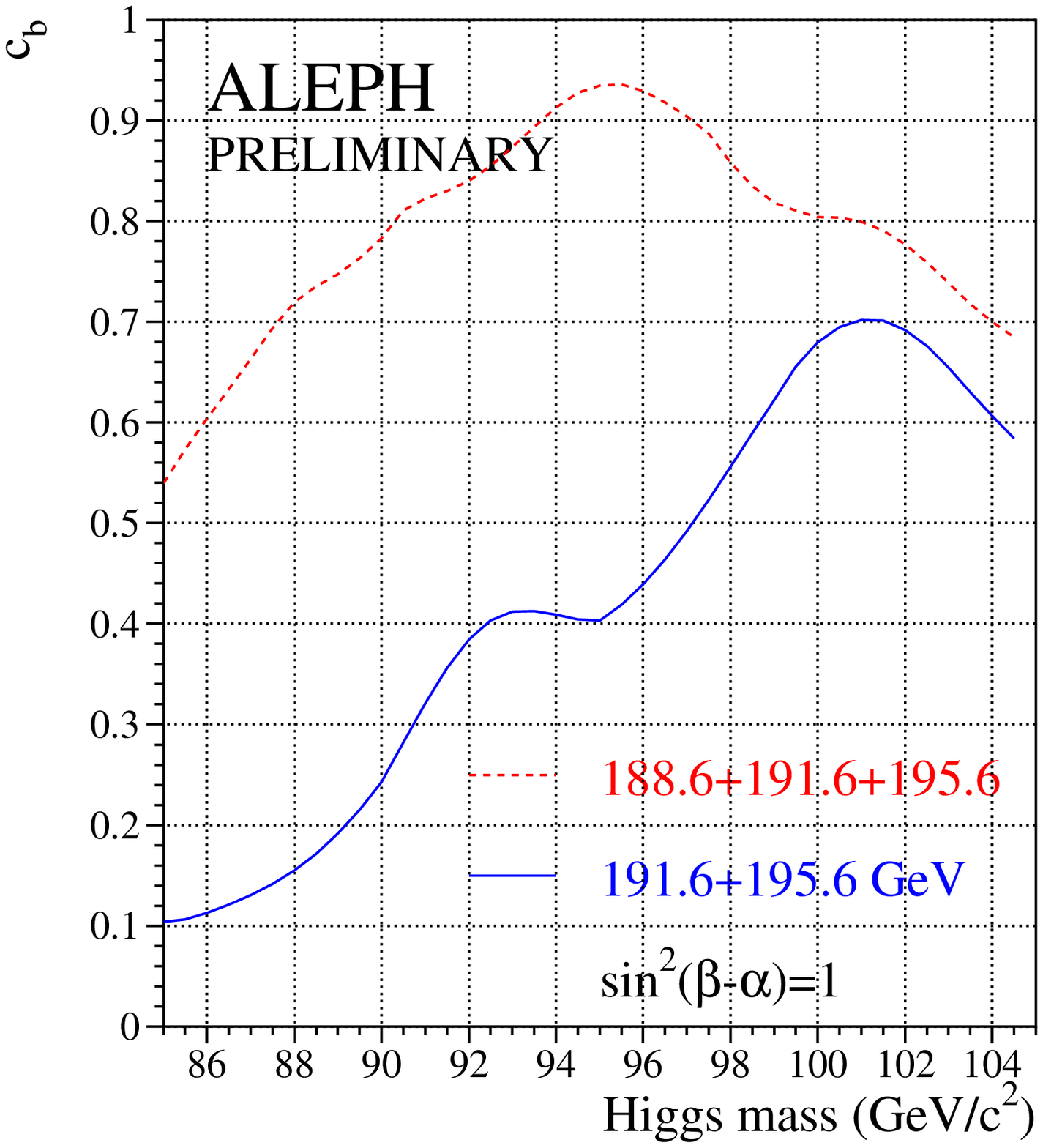}}
\put(70,-5){\epsfxsize100mm\epsfbox{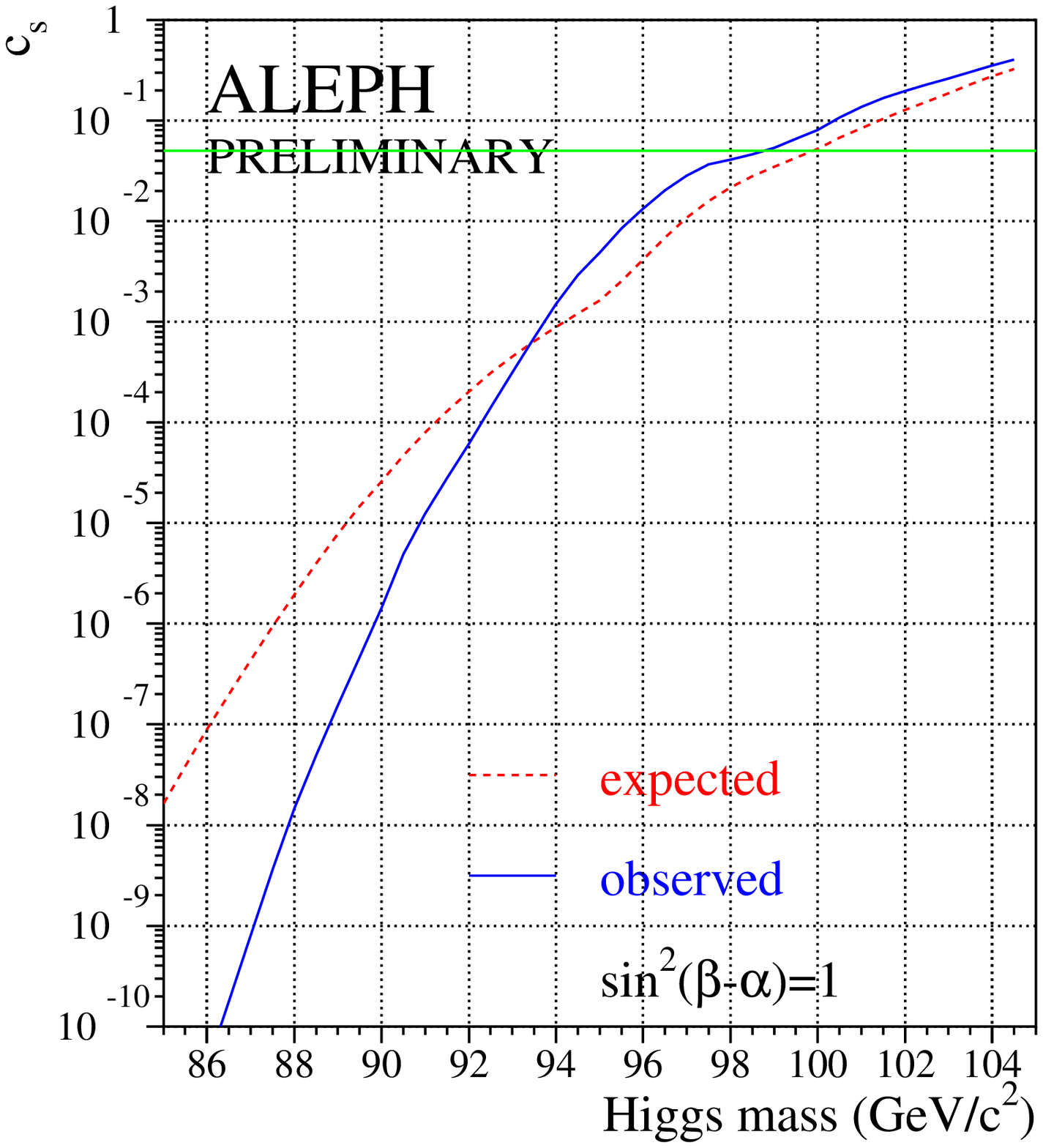}}
\put(10,68){\Large(a)}
\put(95,68){\Large(b)}
\end{picture}
\caption{
  (a) Background-only confidence level curves ($c_b$) as a
  function of the Higgs mass hypothesis for the \ecmA\ and \ecmB~\gev\ 
  data alone (dashed) and combined with the lower energy ALEPH data
  (solid) for \sba=1.  (b) Expected (dashed) and observed (solid)
  confidence level curves for the \SM\ signal hypothesis using all
  ALEPH data up to \ecmB~\gev.
\label{fig:clb}} 
\end{center}
\end{figure}

\begin{figure}
\begin{center}
\begin{picture}(180,180)
\put(-5,0){\epsfxsize160mm\epsfbox{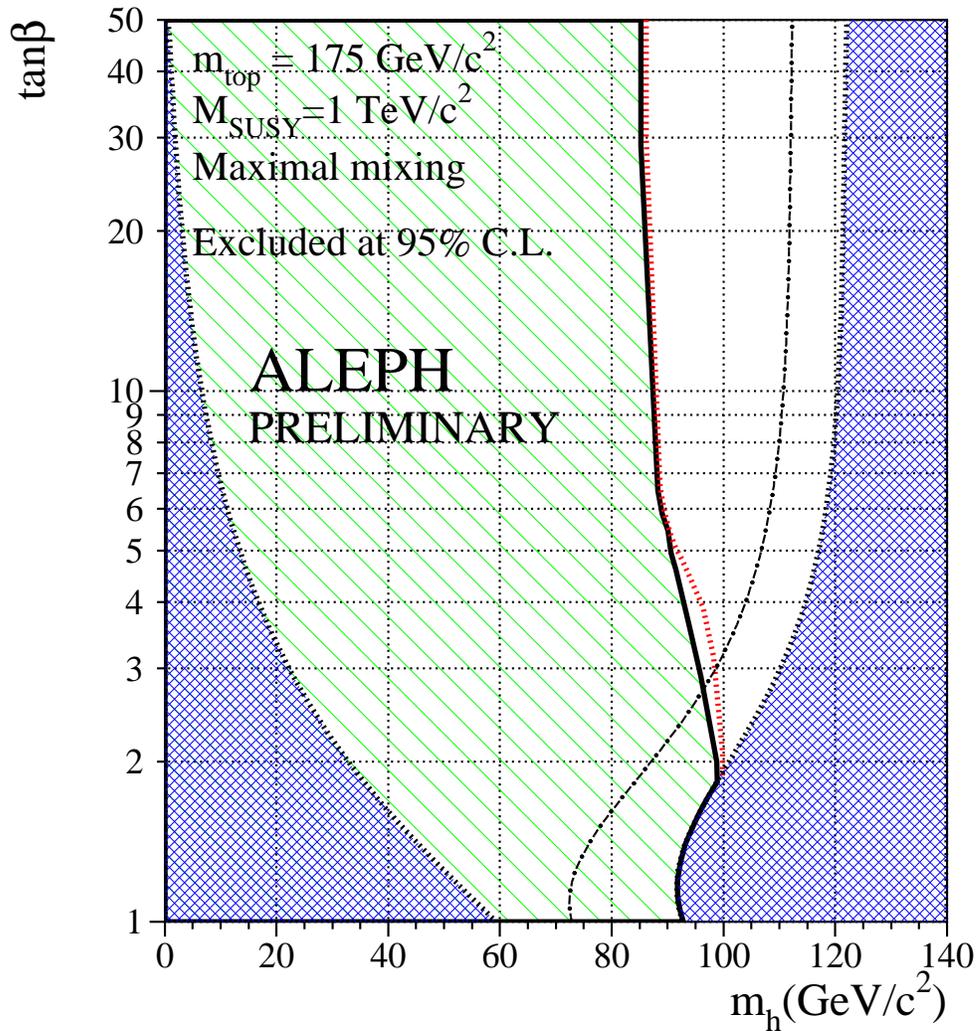}}
\end{picture}
\caption{ 
Expected (dashed) and observed (solid) 95\% confidence level curves  
in the [\mh,\tanb] plane of the MSSM. The dark regions are not allowed
theoretically.  The combined experimentally excluded region is
essentially identical in the case of no stop mixing, for which the
theoretically forbidden region is also indicated (dashed-dotted
curve).
\label{fig:mhtb}} 
\end{center}
\end{figure}

\pagebreak

\end{document}